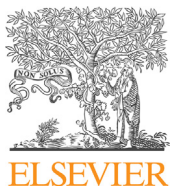
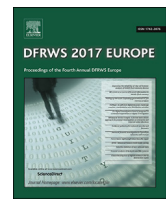

# EviPlant: An efficient digital forensic challenge creation, manipulation and distribution solution

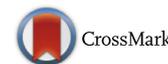

Mark Scanlon[*], Xiaoyu Du, David Lillis

*School of Computer Science, University College Dublin, Ireland*

**ABSTRACT**

Education and training in digital forensics requires a variety of suitable challenge corpora containing realistic features including regular wear-and-tear, background noise, and the actual digital traces to be discovered during investigation. Typically, the creation of these challenges requires overly arduous effort on the part of the educator to ensure their viability. Once created, the challenge image needs to be stored and distributed to a class for practical training. This storage and distribution step requires significant time and resources and may not even be possible in an online/distance learning scenario due to the data sizes involved. As part of this paper, we introduce a more capable methodology and system as an alternative to current approaches. EviPlant is a system designed for the efficient creation, manipulation, storage and distribution of challenges for digital forensics education and training. The system relies on the initial distribution of base disk images, i.e., images containing solely base operating systems. In order to create challenges for students, educators can boot the base system, emulate the desired activity and perform a "diffing" of resultant image and the base image. This diffing process extracts the modified artefacts and associated metadata and stores them in an "evidence package". Evidence packages can be created for different personae, different wear-and-tear, different emulated crimes, etc., and multiple evidence packages can be distributed to students and integrated into the base images. A number of additional applications in digital forensic challenge creation for tool testing and validation, proficiency testing, and malware analysis are also discussed as a result of using EviPlant.

## Introduction

Digital forensic investigators are expected to handle the acquisition of digital evidence from an ever-increasing range of devices, requiring skills originating from a number of different disciplines including law, statistics, governmental policy, psychology, library science, and finance (Palmer et al., 2015). While the spectrum of topics to be covered by any forensics educational programme can be very broad (Nance et al., 2009; Cooper et al., 2010), the focus of this paper is on the sample data used for training and education in digital forensics.

A fundamental issue in forensics and security is that real-world incriminating data is generally unsuitable for educational use (Woods et al., 2011). In order to provide realistic data for training, typically each educational institution creates their own emulated "incriminating" digital data source for investigation, e.g., disk images, network traffic logs, mobile phones device data, etc. Emulating accurate and viable digital evidence for use in the classroom is an extremely arduous task. Currently this process typically requires days or weeks of experts' time (professors, teaching assistants, training personnel) in creating viable digital traces to be discovered during practical investigation training (Yannikos et al., 2014). This project aims to effectively eliminate this wasted time through the development of a methodology for the automated "planting" of digital evidence in a range of standard device images for educational purposes.

The expert effort required for the creation of viable training data comprises of a significant planning phase, a precise execution methodology, and trained personnel to create the resulting evidence. This will typically require manual construction (Moch and Freiling, 2009), e.g., installing a fresh operating system on a physical PC or virtual machine, the installation of common programs (browsers, messaging applications, email clients, file-sharing tools, etc.), and then emulating the necessary user activity can commence. This drawn-out process will likely result in just one usable case study (Garfinkel et al., 2009). There are further delays in the existing process from content creation to sharing the data with a class. Once the viable computer activity has been created, the computer's storage will need to be imaged and distributed. Using current industry standard hard drive investigation tools, such as Encase or Forensic Toolkit (FTK), the time taken to image a entire hard disk is typically in the order of hours.

---

[*] Corresponding author.
E-mail addresses: mark.scanlon@ucd.ie (M. Scanlon), xiaoyu.du@ucdconnect.ie (X. Du), david.lillis@ucd.ie (D. Lillis).

The paper is organised as follows: "Literature review" Section outlines the need for standard corpora for digital forensic education, and reviews existing efforts to provide these. EviPlant is introduced in "EviPlant" Section, where its motivations and functionality are discussed. To date, some initial testing has been performed, which is outlined in "Preliminary testing" Section. Finally, a number of conclusions are drawn in "Concluding remarks" Section and some avenues for continued research are discussed.

*Contribution of this work*

The contribution of this work can be summarised as follows:

- A model consisting of the maintenance of base hard drive images and a methodology for the creation, storage, categorisation, clustering, and indexing of injectable evidence packages.
- A specification for what types of "evidence packages" would be necessary for the creation of realistic emulated machines, e.g., user personae, web browsing histories, regular PC usage patterns, etc.
- The design and prototyping of a solution capable of efficiently creating the necessary evidence packages, alongside a novel evidence planting methodology.

**Literature review**

*The importance of digital forensics education*

In recent years, the increasing proliferation of technology in society has led to an variety of new scenarios in which cybercrimes can be committed. The quantity of crimes that incorporate a digital element has grown also. This explosion in the number of cybercrimes to be investigated has led to pressure on the resources of law enforcement agencies, with backlogs in conducting digital forensic investigations now commonly running into years in many instances (Casey et al., 2009; Lillis et al., 2016).

The combination of the growing quantity of investigations and the rapidly-changing nature of technology has resulted in a strong demand for skilled digital forensic investigators, both in industry and law enforcement, with organisations often experiencing difficulty in filling these positions (Vincze, 2016; Vogel, 2016). Digital Forensics education has a crucial role to play in the provision of new trained investigators, as well as in the continued training and proficiency testing of existing professionals to work in an ever-changing technological landscape.

A looming challenge lies in the fact that training courses directly related to digital forensics are a relatively recent phenomenon. As technology continues to advance, existing qualifications will fast become outdated. This will require substantial continuing professional development for investigators, as well as regular proficiency testing of the type that is common for traditional forensics (Saks and Koehler, 2005).

*Standardised corpora*

The case for standardised corpora is made by Garfinkel et al. (2009), with the primary motivations being reproducibility and education. In terms of education, the authors note that suitable datasets do not occur naturally. There are significant educational, privacy, and legal issues potentially associated with the analysis of students' own systems, systems of students' friends, or hard disks purchased second-hand. The common alternative approach is for educators to spend significant time creating customised data sets.

Al Fahdi et al. (2016) note that the public availability of forensic cases is "very limited". In their work, they make use of two publicly-available cases. The first is "Hunter XP", which provided as a training case for the EnCase digital investigation product. The other was a simulated hacking case that was artificially generated by NIST as part of the CFReDS project.[1] They also gained access to two further cases privately, which required non-disclosure agreements to be signed. This emphasises the level of difficulty associated with obtaining realistic cases for analysis and distribution for educational purposes.

As noted by Woods et al. (2011), the small quantity of available corpora means that solutions to standard datasets are frequently available online, potentially undermining the effectiveness of assessments and proficiency testing. In this scenario, it is desirable that new, unseen challenges be posed to examinees.

*Corpora characteristics*

Woods et al. (2011) describe four ideal characteristics of "realistic" educational corpora (adapted):

1. Answer Keys — these are solutions to the problems posed to students incorporating guidance as to what evidence could be located in which digital artefacts.
2. Realistic Wear and Depth — the sample hard drive images should contain realistic wear patterns, i.e., the hard disk image being investigated should have regular usage surrounding email, web browsing, application installations, file creation and deletion, and downloaded content.
3. Realistic Background Data — a key skill for a digital investigator to gain is the ability to decipher between pertinent and non-pertinent data on a machine. The injection of "incriminating" data should not be obviously the only non-OS/non-application data stored on the disk.
4. Sharing and Redistribution — as a general guideline, hard disk images created for the purposes of education should be made freely available for others to download.

*Current approaches to providing viable disk images*

The problem of providing realistic data for digital forensics education has resulted in a number of techniques being employed by the educator. Moch and Freiling (2009) outline three existing approaches to the creation or acquisition of digital forensic datasets or viable disk images:

- Perhaps the most widespread method is the *manual creation* of disk images. Here, an instructor creates a disk image that contains specific evidence for students to find. This has the advantage that the precise evidence is known to the instructor and can be used for evaluation purposes. Additionally, there is no requirement to wait for interesting activity to occur in a natural setting, as the instructor is free to perform/emulate any actions that are desired. However, creating these images is a very time-consuming task, particularly given the requirement to ideally provide realistic wear and depth.

---

[1] http://www.cfreds.nist.gov.

- A *honeypot* involves connecting a computer to a network with the express intention of it being attacked and compromised (Provos and Holz, 2007). By recording the activities of attackers, interesting disk images can be created. However, the majority of attacks are automated, and the quantity of images that feature manual attacks for students to study is low. Due to the low quantity of interesting examples available, analysis results can frequently be found online. From a suitability standpoint, the required analysis of these honeypot generated challenges is often at too high a difficulty level for many learners (Woods et al., 2011).
- A fruitful source of realistic data is *second-hand hard disks.* This approach results in valuable data on naturally occurring phenomena on disks, as the disks have typically been in use by a real user over a longer period of time than an instructor can dedicate to the manual creation of an image (Freiling et al., 2008). Pre-used hard disks are the source of the Real Data Corpus, assembled over a number of years by Garfinkel (2007, 2012). This forms part of a 30TB collection of research corpora, which also includes items such as network packet traces, known malware and a million document corpus gathered from the *.gov TLD.

One drawback of this approach is that it does not include materials relating to real crimes that could be used for training purposes (Baggili and Breitinger, 2015). Additionally, the use of data belonging to real users raises a number of legal data protection issues. As the data is generated by real users, privacy law (which greatly varies by jurisdiction) must be taken into account, particularly when redistributing images. Images may also contain copyrighted materials (including the operating system and software) or illegal files.

Some attempts have been made to create tools that automate the generation of disk images for education purposes. This approach attempts to leverage the advantages of manual creation, while expediting the process.

Forensig[2] is one such automation tool (Moch and Freiling, 2009, 2011). This allows instructors to write script files that are executed by the system to simulate certain user behaviour. The output of the process is an image file for students to analyse, and a "ground truth" that can later be used to evaluate their performance. The scripting language used by Forensig[2] allows the instructor to add randomness to the build process, which facilitates the generation of distinct images that nonetheless contain the same investigative challenge. In order to maintain the reproducibility of challenges, a two-pass process is used, whereby the second pass is based on an intermediary, deterministic input script. Any random decisions incorporated in the original script are taken during the first pass.

Another automated image creation tool is ForGe (Visti et al., 2015). This also allows a user to set up a scenario, including random elements, which results in the generation of an NTFS disk image. In generating the image, it utilises a number of data hiding techniques, e.g. placing data in file slack or unallocated space. Unlike Forensig,[2] which is script-driven, ForGe provides a graphical interface to facilitate the creation of images.

**EviPlant**

This paper introduces EviPlant as a more efficient alternative for the creation, manipulation, storage, and distribution of digital forensic challenges to classes of students. The fundamental premise of EviPlant is that a base disk image file can be downloaded by students once, and challenges can then be distributed as much smaller "evidence packages". These evidence packages can then be integrated with the base image to create the disk image for analysis. This technique addresses a significant drawback in using separate full image files for each challenge: namely that these images may be in the order of tens or hundreds of gigabytes in size. For students to download such large images during each class can be very time-consuming, and can burden even high bandwidth networks and servers. In an online/distance learning scenario, this problem is further compounded by students' own internet connection speeds, which can vary widely, and can make distribution infeasible or effectively impossible. This also avoids the requirement on students to dedicate large amounts of disk space to storing multiple disk images that they have been asked to analyse.

EviPlant consists of two core components: a *diffing tool* and an *injection tool*. To create the challenge, an instructor can boot the standard base image in a virtual machine, and emulate the criminal behaviour that the students are tasked with detecting. The diffing tool then compares the base image with the instructor's modified image, resulting in an evidence package that includes all files and other digital artefacts that have been created or altered, along with their associated metadata. This package can then be distributed to students, who then use the injection tool to "plant" this evidence on their base images, as can be seen in Fig. 1.

Using this tool, it is possible to develop a methodology and technical standard for the automated injection of digital evidence in a range of device images for educational purposes. It is hoped that this will help to overcome some of the issues encountered by Garfinkel (2012) in managing a large number of complete disk images.

*Design considerations for EviPlant*

In order for EviPlant to provide the same functionality available using current techniques and to also provide additional features, a number of design considerations emerge including:

- *Ease of Creation* – Switching from the manual approach to creating viable datasets for classes should not require significant effort on the part of the educator. The process of creating and curating the challenges should be straightforward.
- *Efficient Distribution* – One of the problems with each of the aforementioned approaches is that the distribution of the resultant disk images takes significant time. Courses are limited in the range of images distributed to a class due to the resourcing and time required for their distribution. This issue is compounded in an online/distance learning situation. The EviPlant solution should greatly reduce the file sizes need for distribution to classes. This will enable a broader range of investigation types to be covered by a course and provide students with viable, formative practice problems.
- *Efficient Injection* – The planting of the digital artefacts and associated metadata should be as efficient as possible. As one of the main advantages of EviPlant is facilitating a greater variety of challenges by the learner, the impact of evidence injection should be as minimal as possible.
- *Operating System Compatibility* – The operating system used by the educator or the students should not be of concern for the usage of the system. Likewise, the educator should be permitted to choose the "suspect" machine's operating system according to educational needs.
- *Mobile Compatibility* – Mobile device forensic training should be possible using the system in a similar manner to the desktop/server paradigm. The system should be compatible with modern mobile operating systems, e.g., iOS, Android, Windows Mobile, etc.

---
[2] https://moodle.org/.

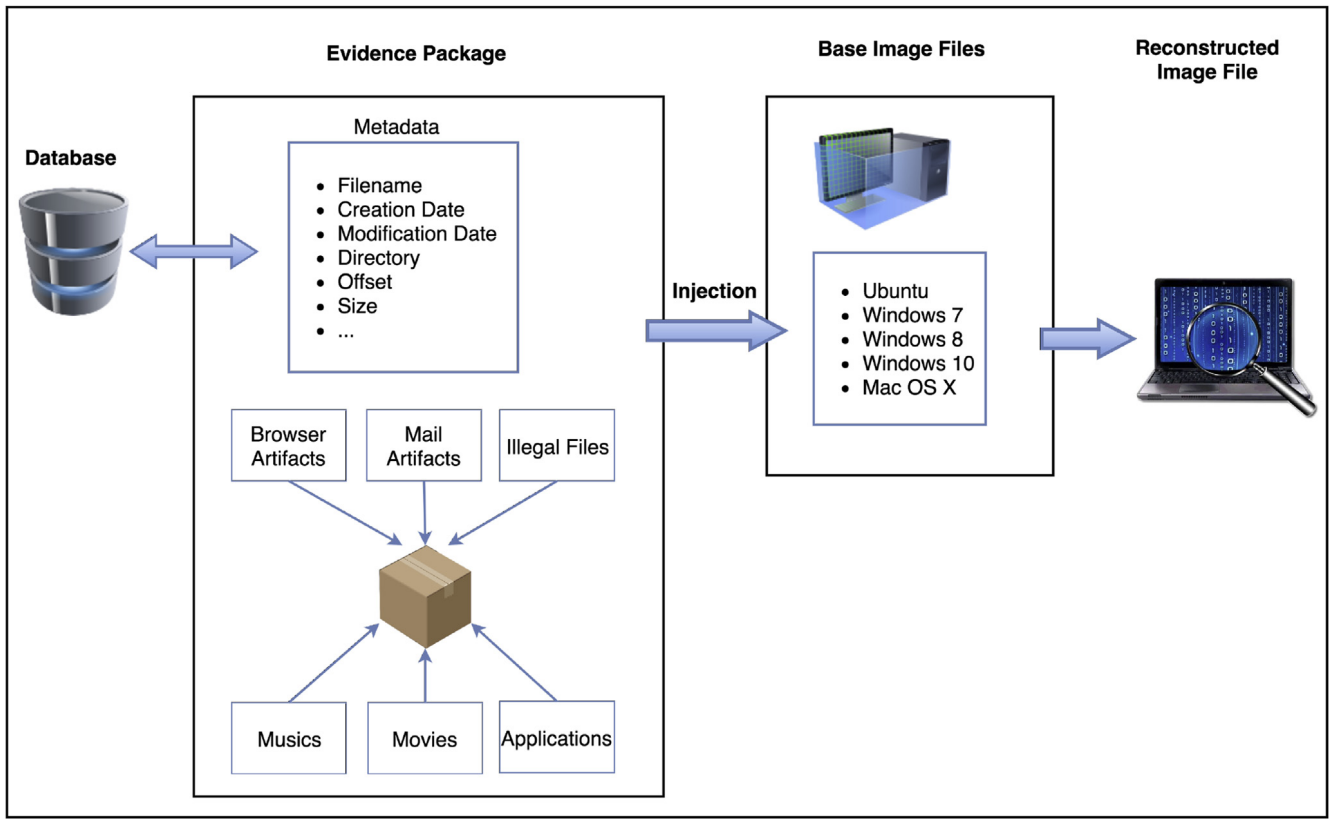

**Fig. 1.** Overview of methodology.

- "Story Mode" – To eliminate the burden on the educator, the combination of numerous evidence packages (each containing different usage patterns) into the one challenge should be possible. Assuming a sufficient number of evidence packages have been created, "stories" should be constructable/manipulable from a variety of sources to create realistic problem sets.
- Answer Sets – As with any educational system, the images generated from EviPlant should be usable from both a formative and summative perspective. The automated generation of suitable answer sets will contribute to the reduction of effort on behalf of the educator in the creating and grading of assignments. Due to the curated nature of the base image and specific evidence packages (both wear-and-tear and pertinent), the generation of answer sets for the challenges produced by the system should be straightforward.

*Evidence packages*

The fundamental building blocks for EviPlant are evidence packages. Evidence packages contain all the digital artefacts (files, file fragments, slackspace, etc.) and associated metadata created during the emulation of the crime. The artefacts contained within evidence packages fall into two categories, which combine to capture all the necessary data modified from the base image after performing a specific task or set of tasks:

1. "Black Box" Artefacts – These are simple artefacts encapsulating a modification from the base image. In order to use these evidence packages in a classroom deployment scenario, nothing need necessarily be understood of their construction in order to inject them into the base images. The answer set for these challenges is created by the educator (effectively the script they followed while performing the actions to be discovered).
2. "Reversed" Artefacts – The artefacts and metadata contained within these package are understood in their entirety – effectively reverse engineered. As a result, their manipulation is possible, e.g., SQLite databases for Internet browsing history, VoIP application call logs, etc. In the scenario of using multiple evidence packages for a single challenge, packages with overlapping artefacts must be manipulable to ensure that the traces from each packages are integrated into the final disk image.

In a realistic usage scenario, a catalogue of evidence packages would be created containing different user profiles/personae, different application usage patterns, different browsing history, different background noise, etc. These different evidence sources are nestable, with an conflict (conflicting artefacts or metadata) between later evidence packages and subsequent evidence packages being resolved by the former overwriting the latter. Seeing as many of these wear-and-tear and background data tasks are already being performed by educators who create their own challenges, the sharing of these background packages between educators would be of mutual benefit. Due to their minimal file sizes, the sharing of these packages would require minimal data transfer.

*Evidence creation methodology*

The current approach requires the maintenance of large collections of complete disk images in order to provide a variety of example challenges to a class. Using EviPlant, the approach changes to creating and curating a variety of evidence packages. The creation of these evidence packages relies on the comparison or "diffing" of two disk images ("diffing" in this context stemming from the diff tool in *nix systems (Hunt and McIlroy, 1976)). The diffing tool is provided with the base image used to emulate the specific user activity and the subsequent modified image containing all traces of the "crime", wear-and-tear, or persona emulated. The tool scans through the modified image and extracts all modified and newly created artefacts (and their associated metadata) into an evidence package, i.e., eliminating all artefacts from the modified image also present on the base image.

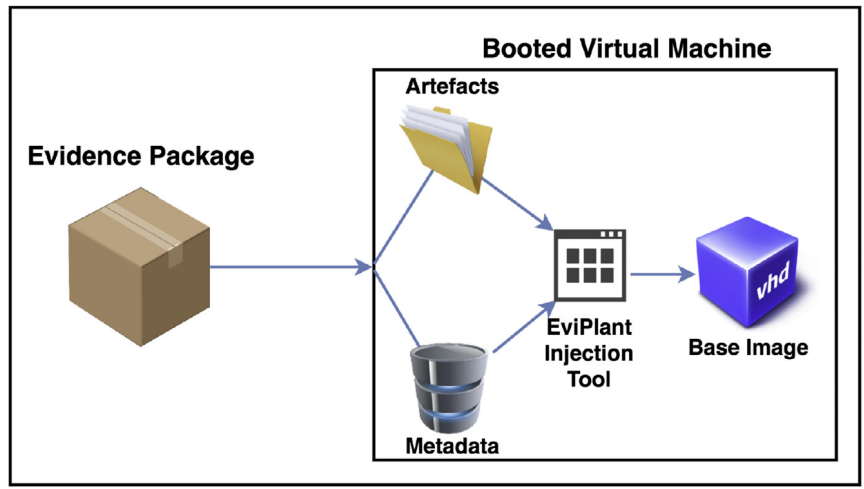

Fig. 2. Logical evidence planting.

In this manner, evidence packages are capable of being created to capture small events, e.g., a boot cycle of the operating system, or large events, e.g., a complex usage pattern to build a complete emulated user persona over an extended period of time.

*Distribution*

In terms of distribution of the challenges to students, due to the utilisation of evidence packages greatly reducing the file size required for distribution to each member of a class, regular file transfer services become more viable. For example, the uploading of the requisite evidence packages to virtual learning environments such as Moodle[2] or Blackboard[3] becomes more feasible. Previously, the uploading of complete disk images may not have been possible due to file size and/or bandwidth constraints. If the need arose to distribute larger, more complex challenges (multiple gigabytes) to students, peer-to-peer file distribution methods, such as BitTorrent Sync (Farina et al., 2014), might be the most performant by sharing the distribution workload among the class themselves.

*Evidence planting*

The model for evidence injection centres around the initial distribution of a base image (or collection of base images if necessary) to each student in a class. The premise of EviPlant is that the same base image will be used multiple times throughout the course, thus reducing the overall volume of information to be shared. Each base image consists of a bare installation of any desired operating system and would be cloned before evidence planting would commence, in order to facilitate easy reuse. In an online/distance learning scenario, remote learners could either download the base image first (with this base image being useful for many exercises) or they could bring their own OS (but would have to ensure it matches that of the educator).

This model has the additional advantage that it avoids questions of copyright infringement that may arise when distributing entire disk images that contain an operating system and other software. Educational institutions typically maintain their own site licenses for software, which means that educators and students alike may begin from a fully-licensed base image. Evidence packages can then be shared between institutions more freely.

For the injection of "black box" packages, the evidence packages (containing all artefacts and associated metadata) are downloaded from a centralised server. These artefacts are placed in the base image, overwriting any overlapping data between the base image and the evidence package. For the "reversed" packages, manipulation of the metadata is possible before injection. For example, the internet browsing history or VoIP call logs can be updated to have occurred at different time than the creation of the package.

There are two options available for the evidence injection process itself:

- Logical Data Injection – This option requires the booting of the disk image in a virtualised environment and the execution of the injection tool natively in the guest operating system, as can be seen in Fig. 2. The injection tool downloads the evidence package(s) from a centralised server and processes the artefacts one-by-one. Any existing files in the operating system are overwritten by those in the evidence package.

This method provides the benefit of the elimination of any configuration issues with the injection tool running on students' machines. In fact, if this is the injection methodology of choice, the injection tool would come pre-installed on the base images before initial distribution and the student can select the relevant assignment during boot. Of course, this logical injection method will leave traces of the tool itself on the disk image, but this is unlikely to be any concern in a educational context.

- Physical Data Injection – This option involves the modification of the underlying blocks of the base disk image, as can be seen in Fig. 3. The injection tool runs on the students' native operating system and works directly with the base image file. Similarly to the previous scenario, the tool would still download the evidence package(s) from the centralised server. The data would be written to the disk image at their corresponding offsets, ignoring the make-up of the existing file system. Physical injection is the less intrusive method to the target disk image and is necessary for verifiable disk image reconstruction. No trace of the injection tool would be present on the resultant image.

[3] http://www.blackboard.com/.

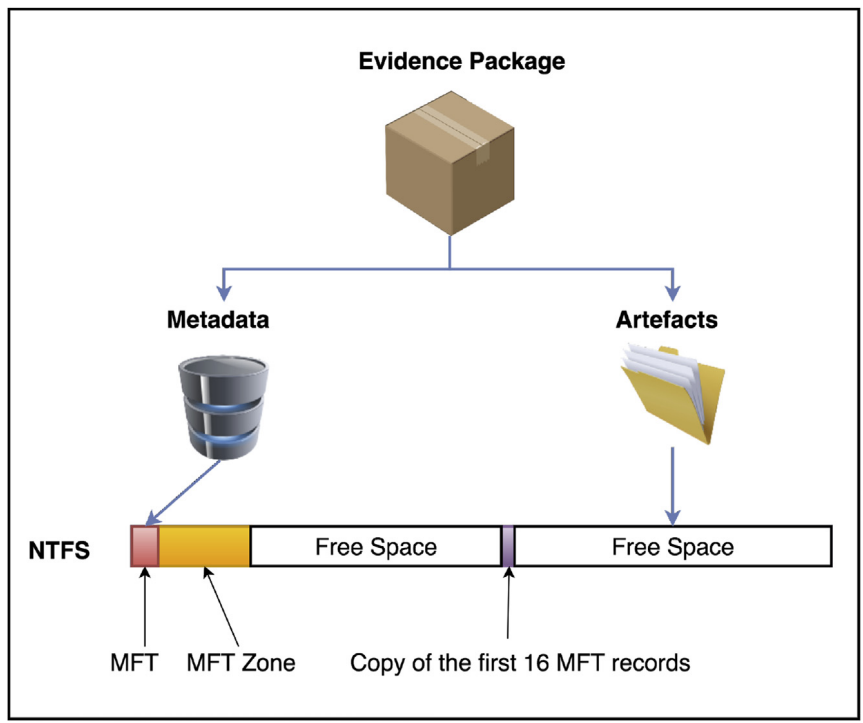

**Fig. 3.** Physical evidence planting.

*Additional educational benefits enabled through EviPlant*

The potential impact of the proposed system extends beyond the time saved by expert educators in the creation and distribution of realistic content for their classes and enables a number of additional educational benefits to enhance learning:

- Helps to Eliminate Plagiarism – Custom generated, practical digital forensic challenge exercises eliminate the possibility of students engaging in plagiarism of results for known, freely available corpora. After building a sufficiently large corpus of evidence packages, it is possible that a unique disk image could be automatically created per student with each offering their own challenges for the students while achieving the learning outcomes of the current topic.
- Automated Practice Exercises – The ability to create exercises for students on-the-fly will allow students to practise their skills on many different exercises as opposed to being limited to the limited number of disk images made available to them.
- Timeline Emulation – Due to the minimal footprint of each evidence package, frequent diffing becomes feasible for a running disk image. As a result, studying the evolution of a case study becomes possible with a different evidence package being created at each event milestone.
- Assessment – The creation of a different challenge for each student in a class can enable laboratory based practical assessment. Multiple students could take this test simultaneously in the same room as a bespoke challenge could be set for each.

Outside of a educational usage scenario, EviPlant could also be used for the curation of evidence packages and challenges suitable for:

- Proficiency Testing – In order to provide viable proficiency testing for digital forensic investigators, a comprehensive suite of challenges must be available. Building a sufficient catalogue of evidence packages allows the quick creation of challenges for proficiency testing containing numerous permutations and combinations of the catalogue. In the scenario of multiple investigators being tested simultaneously at the same location, each investigator could be tested for the same skills while working on different challenges.
- Forensic Software Testing and Validation – The testing and validation of forensic software is an important issue (Beebe, 2009; Guo et al., 2009). Often, forensic software is tested against common datasets, which are also generally available to the software developers. Another potential use for EviPlant is to create corpora on which forensic software can be tested, and evaluated with regard to its success rate in identifying the simulated criminal behaviour.
- Point-in-Time Reconstruction – Using a similar technique as that outlined above for timeline reconstruction training, high frequency package creation would enable the real-time monitoring and reconstruction of the device state at any point necessary. In order to increase the frequency of the packages, they would be created against the previous snapshot as opposed to the original base image. This would then enable point-in-time reconstruction by sequentially integrating the evidence packages in order of their acquisition time – similar to the incremental backup approach used by database administrators.
- Malware Analysis – By intentionally installing malware on a test image, the diffing tool can be repurposed to provide evidence packages that documents the malware's lifecycle on a target system. After the fact, the system can be reconstructed for the analysis of the malware at any stage in its lifecycle.
- International Collaboration – Current international collaboration techniques typically involve the shipping of seized devices to the requesting agency. Using the proposed approach in combination with a data deduplication system, such as that proposed by Scanlon (2016), can greatly expedite the international transmission of evidence.

**Preliminary testing**

To demonstrate the viability of the proposed system, the two main components of the system (namely the diffing tool and the injection tool) were developed in Python using the pytsk[4] library for disk image analysis. pytsk is a python wrapper for The Sleuth Kit.[5] The Sleuth Kit provides a wealth of file system compatibility including NTFS, FAT, ExFAT, UFS 1, UFS 2, EXT2, EXT3, EXT4, HFS, ISO 9660, and YAFFS2. For testing purposes, a Windows 10 virtual machine was created and used as the base image. For each test, the base image was cloned, booted and user activity was emulated on the machine.

In terms of the testing of the diffing tool, a variety of usage patterns were tested ranging from a single boot cycle of the virtual machine, to an extended session involving internet browsing, application installation, file downloading, multiple boot cycles, etc. In each scenario, the diffing tool discovered all of the modified artefacts relating to recorded usage, including a number of operating system files that were modified during the regular usage of the virtual machine (e.g., $MFT, pagefile.sys, etc.). These artefacts and associated metadata were output into an evidence package.

To assess the injection methodology, the base operating system was booted and the injection tool was run locally on the live machine to perform a logical injection of the artefacts, as described in "Evidence planting" Section. The injection tool took each individual artefact and added it to the virtual disk. In the eventuality of conflicting artefacts, the version from the base image was overwritten by that from the evidence package. To demonstrate the viability of the resultant generated disk images and to confirm the injection of the necessary artefacts, these images were subsequently analysed using EnCase and, unsurprisingly, the pertinent planted evidence was identifiable and recoverable.

**Concluding remarks**

The solution presented as part of this paper makes the creation of digital forensic challenges easier for the educator. More digital forensic challenges are capable of being stored in the same disk capacity than if entire disk images were used. This reduction in required storage can facilitate more challenges being given to students to enhance their learning, i.e., the distribution time and local storage required on students' machines are both reduced. While EviPlant is focused on enhancing digital forensics education, the approach outlined above for the creation and manipulation of hard disk images can also be applied to complimentary fields, e.g., digital forensic tool testing and validation, virtual machine/cloud instance monitoring, point-in-time reconstruction, etc.

*Future work*

The EviPlant system outlined as part of this paper is currently a functional prototype. As with any system at the prototype stage, there are a number of desired features that will be developed upon in the future including:

---

[4] https://github.com/py4n6/pytsk.
[5] http://www.sleuthkit.org/.

- Manipulation of Evidence Packages – If components of the evidence package are understood, e.g., SQLite database, file traces, etc., the manipulation of these components should be possible.
- Collision Resolution – Currently, the integration of evidence packages into a base image overwrites any artefacts that were previously contained in the image. Multiple evidence packages can be integrated in succession, but any collisions, i.e., the same artefact contained in both packages, will result in that artefact being overwritten. To provide intelligent collision resolution (for both "black box" and "reversed" packages) would greatly expand the complexity of challenges creatable and the broader usefulness of the tool.
- Physical Injection – Adding forensic artefacts to the disk without initially booting the OS is desirable as any traces of the injection tool itself executing on the disk image would be eliminated. While not an absolute requirement for educational purposes, for a number of the complimentary applications of the technology this may prove necessary.
- Comprehensive Evaluation – Implementation and evaluation of a real-world deployment of the tool in both face-to-face and online/distance classroom settings.